\documentclass[superscriptaddress,showpacs,aps,twocolumn,prb,floatfix]{revtex4}
\usepackage{amssymb}
\usepackage{amsmath}
\usepackage[dvips]{graphicx}
\usepackage{bm}

\begin{document}
\title{Replicas of the Kondo peak due to electron-vibration interaction in molecular transport properties}

\author{P. Roura-Bas}
\affiliation{Dpto de F\'{\i}sica, Centro At\'{o}mico Constituyentes, Comisi\'{o}n
Nacional de Energ\'{\i}a At\'{o}mica, Buenos Aires, Argentina}

\author{L. Tosi}
\affiliation{Centro At\'{o}mico Bariloche and Instituto Balseiro, Comisi\'{o}n Nacional
de Energ\'{\i}a At\'{o}mica, 8400 Bariloche, Argentina}

\author{A. A. Aligia}
\affiliation{Centro At\'{o}mico Bariloche and Instituto Balseiro, Comisi\'{o}n Nacional
de Energ\'{\i}a At\'{o}mica, 8400 Bariloche, Argentina}

\begin{abstract}
The low temperature properties of single level molecular quantum dots
including both, electron-electron and electron-vibration interactions, are
theoretically investigated. The calculated differential conductance in the
Kondo regime exhibits not only the zero bias anomaly but also side peaks
located at bias voltages which coincide with multiples of the energy of
vibronic mode $V \sim \hbar\Omega/e$. We obtain that the evolution with
temperature of the two main satellite conductance peaks follows the
corresponding one of the Kondo peak when $\hbar\Omega \gg k_B T_K$, being $
T_K$ the Kondo temperature, in agreement with recent transport measurements
in molecular junctions. However, we find that this is no longer valid when $
\hbar\Omega$ is of the order of a few times $k_B T_K$.
\end{abstract}


\maketitle

\section{Introduction}

The Kondo effect, originally discovered in metals containing magnetic impurities,\cite{hewson,kondo} 
is also observed in transport measurements through semiconducting \cite{gold,cro,gold2,wiel,grobis,kreti,ama} 
and molecular \cite{liang,kuba,yu,leuen,parks,roch,scott,parks2,serge,vincent} quantum dots (QDs)
in which the QD acts as a single magnetic impurity.
While the semiconducting QDs are characterized by the tunability of its parameters and 
have served as platforms to study the one- and two-channel,\cite{gold,gold2,grobis,2ch} as well as 
the SU(4) Kondo effect,\cite{su4} the molecular QDs (MQDs) have
allowed researchers to investigate, among different phenomena, 
the underscreened Kondo effect for spin $S > 1/2$ \cite{parks,roch,serge} 
and quantum phase transitions 
driven by stretching \cite{parks,corna} or gate voltage.\cite{roch,serge} Remarkably, 
MQDs incorporate the effect of phonons. 
Molecular vibration signatures have been observed in conductance measurements through a variety of molecules, 
such as H$_2$,\cite{hidrog-molec} and C$_{140}$.\cite{c-140}
Furthermore,
experiments performed in the Kondo regime have revealed the presence of satellite peaks at finite bias 
which emerge together with the zero-bias anomaly in the differential conductance as a consequence of the 
interplay between the electron-vibration interaction and the many-body Kondo state.
\cite{parks, park, zhitenev, yu2, fernandez, rak}

Recently, Rakhmilevitch \textit{et al.} reported on transport measurements through a 
copper-phthalocyanine (CuPc) molecule connected to two silver contacts in a break-junction setup.\cite{rak}  
Their work focuses on the evolution with temperature of the side peaks observed in the differential conductance. 
The authors find that the maximum conductance of the side peaks increases when the temperature 
is lowered following the same dependence as the zero-bias Kondo peak. 
Specifically (since CuPc possess a spin-$1/2$,\cite{rak,mugarza}) the 
intensity of both, the zero-bias anomaly and the satellite peaks, can be fitted with
the same empirical expression for the temperature-dependence of the equilibrium 
conductance of a spin-$1/2$ impurity, $G(T)$, which follows closely results obtained using 
the numerical renormalization group
\begin{equation}
G(T,V=0)=\frac{G_{s}}{\left[ 1+\left( 2^{1/s}-1\right) \left( T/T_{K}\right)
^{2}\right] ^{s}},  \label{ley_empirica}
\end{equation}
where $s=0.22$,\cite{gold, G_E} $G_s$ is the conductance at temperature $T=0$, and 
the Kondo temperature $T_K$ is the only adjustable parameter. 
Outstandingly, the fitted Kondo scale for the satellite peaks 
located at bias voltage $ eV \sim \pm \hbar\Omega\sim \pm 21\,$me$V$, agrees with the one obtained for 
the Kondo peak (21\,K $< T_K <$ 25\,K). 
This is an interesting result for two main reasons: first, it is not evident that the empirical expression 
Eq. (\ref{ley_empirica}), that correctly gives
the universal temperature dependence of the conductance without vibrational modes, can be 
applied when the Kondo phenomena is assisted by phonons. 
More surprising is the fact that Eq. (\ref{ley_empirica}) still works out of equilibrium, for fitting 
the conductance at finite bias voltages of the order of $\vert V\vert\sim \hbar\Omega/e$. 
While the satellite peaks have been studied theoretically before,\cite{paaske,roura_1} their dependence on 
temperature has not been analyzed.

Motivated by the experiment of Rakhmilevitch \textit{et al.}, in this article we investigate 
theoretically the low-temperature transport properties of a single level molecular quantum dot including both, 
electron-electron and electron-vibration interactions. 
In agreement with the experimental results, we obtain that the conductance of the two main satellite peaks 
follows the same temperature evolution as
the corresponding Kondo peak, which means that these peaks are also a manifestation of the Kondo effect, 
when $\hbar\Omega \gg \textit{k}_B T_K$, which is in fact the regime of the experiment. 
On the other hand, the statement is no longer
valid when both energy scales are similar, $\hbar \Omega \approx 3\mathit{k}
_{B}T_{K}$. For $\hbar \Omega \leq \mathit{k}_{B}T_{K}$, the satellite peaks
merge with the Kondo peak.\cite{hm}

\section{Model and formalism}

We model the MQD with the Anderson-Holstein Hamiltonian 
\cite{paaske,roura_1,hm,lili,corna2,zitko,mon} in which a spin-$1/2$ doublet of
energy $E_{d}$ is connected to two metallic reservoirs and also coupled to a
phonon mode of frequency $\Omega $, through the electron-phonon interaction 
$\lambda $. The Hamiltonian is 
\begin{eqnarray}
H &=&\left[ E_{d}+\lambda (a^{\dagger }+a)\right] n_{d}+Un_{d\uparrow
}n_{d\downarrow }+\sum_{\nu k\sigma }\epsilon _{k}^{\nu }c_{\nu k\sigma
}^{\dagger }c_{\nu k\sigma }  \notag \\
&&+\sum_{\nu k\sigma }(V_{k}^{\nu }d_{\sigma }^{\dagger }c_{\nu k\sigma }+
\mathrm{H.c}.)+\Omega a^{\dagger }a,  \label{ham}
\end{eqnarray}
where $n_{d}=\sum_{\sigma }n_{d\sigma }$, $n_{d\sigma }=d_{\sigma }^{\dagger
}d_{\sigma }$, $d_{\sigma }^{\dagger }$ creates an electron with spin 
$\sigma $ at the relevant state of a molecule (or quantum dot), $a^{\dagger }$ 
creates the Holstein phonon mode,  $c_{\nu k\sigma }^{\dagger }$ creates a
conduction electron at the left ($\nu =L$) or right ($\nu =R$) lead, and 
$V_{k}^{\nu }$ describe the hopping elements between the leads and the
molecular state. We take the limit of very large Coulomb repulsion 
$U\rightarrow \infty $. 

We use the non-crossing approximation (NCA) in its non-equilibrium extension.
\cite{win,roura_1} The out of equilibrium NCA approach has proved to be a
very valuable technique for calculating the differential conductance through
a variety of systems including two-level QD's and C$_{60}$ molecules
displaying a quantum phase transition,\cite{serge,roura_2, tosi_1} among
others. Furthermore, it is specially suitable for describing satellite peaks
away from the zero bias voltage \cite{tosi_2,NFL,st} and captures the
universal behavior in the equilibrium conductance given by 
Eq. (\ref{ley_empirica}).\cite{roura_3,note3} The application of the NCA to the model
and its limitations for large $\lambda $ are described in detail in Ref. 
\onlinecite{roura_1}

The current through the molecule is calculated using the exact expression 
\cite{meir,win} 
\begin{equation}
I(V)=\frac{4\pi e}{\hbar }\frac{\Gamma _{L}\Gamma _{R}}{\Gamma _{L}+\Gamma
_{R}}\int d\omega \rho (\omega )\left( f(\omega -\mu _{L})-f(\omega -\mu
_{R})\right) ,  \label{current}
\end{equation}
where $\Gamma _{\nu }=2\pi \sum_{k}|V_{k}^{\nu }|^{2}\delta (\omega
-\epsilon _{k}^{\nu })$ (assumed independent of energy) is the coupling of
the molecule to the lead $\nu $, $f(\omega )$ is the Fermi distribution
and the spectral function of the molecule is given by $\rho (\omega )$,
which we calculate using the NCA. The right and left chemical potentials 
$\mu _{\nu }$ of the metallic contacts are proportional to the bias voltage.
For simplicity we assume a symmetric voltage drop $\mu _{L}=-\mu _{R}=eV/2$.
The results are not affected by this assumption.

The reduced intensity of the zero-bias peak in the experiment suggests that there is a large asymmetry 
between the two tunneling couplings $\Gamma_{\nu}$. This is usually the case in MQDs. 
We choose $\Gamma_R\approx50\Gamma_L$ (similar values do not affect our conclusions). 
We note that for ratios of the couplings larger than 10 (i. e. highly asymmetric devices), 
the differential conductance $G=dI/dV$ at bias voltage  $\vert V\vert \lesssim k_B T_K/e$ 
reproduces the equilibrium spectral density of the Kondo resonance.\cite{capac}

While the value of the vibration frequency is well defined by the position of the satellite peaks in the conductance measurements ($\hbar\Omega \sim 21$\,meV), 
neither the value of the total coupling $\Gamma_R+\Gamma_L$ nor the energy position of the Kondo active molecular level $E_d$ are clearly determined. Therefore, we analyzed several values of both energies to confirm that our conclusions remain the same.
The estimated value for the electron-vibration interaction $\lambda$, 
that corresponds to a breathing mode of the CuPc molecule, is found from ab-initio calculations to be near $\lambda_0=6$\,meV. Since the Kondo scale strongly 
depends on the electron-phonon coupling,\cite{roura_1} in the present work, we tested several values of $\lambda$, 
from $6$ to $12$\,meV. We verified that our analysis regarding the ratio 
$k_B T_K/\hbar\Omega$, holds for the whole set of $\lambda$'s studied.

\section{The main Kondo peak}

\begin{figure}[t]
\includegraphics[clip,width=6.0cm]{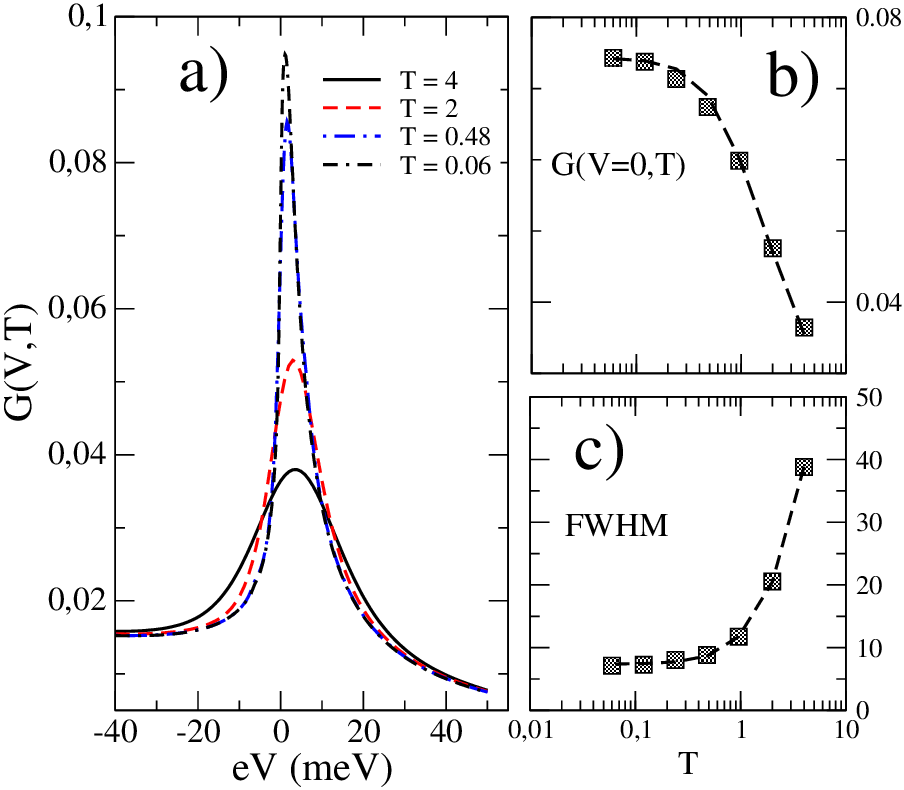}
\caption{(Color online) (a) Differential conductance as a function of bias 
voltage for several temperatures $T$. Parameters in meV:
$\Gamma_R+\Gamma_L=80$, $E_d=-90$, $\lambda=0$. 
(b) Equilibrium conductance
$G(V=0,T)$ as a function of temperature (squares) and the corresponding scaling using Eq. (\ref{ley_empirica}) 
(dashed line) being $T_K = 43 \pm 1$\,K. 
(c) FWHM of the zero bias anomaly as a function of temperature (squares) and the 
fitting function (see text) with $\alpha=9.5 \pm 0.5$ and $T_K = 43 \pm 1$\,K (dashed line).}
\label{fig-1}
\end{figure}

We start by a brief review of the temperature evolution of the Kondo peak in Fig. \ref{fig-1}(a) 
and the scaling law of the zero-bias conductance with Eq. (\ref{ley_empirica}) in Fig. \ref{fig-1}(b)
in absence of the electron-phonon interaction. The asymmetry in $G(V)$ is due to the
asymmetric $\Gamma _{\nu }$.\cite{capac}
For the parameters of the figure, the Kondo scale resulting from the fit using Eq. (\ref{ley_empirica}), 
$T_K = 43 \pm 1$\,K, remarkably agrees with the corresponding one in Fig. \ref{fig-1}(c) extracted 
from a fitting of the full width at half maximum (FWHM) of the zero-bias anomaly using the 
expression \cite{tsukahara,comment_2} FWHM$ = \frac{1}{e}\sqrt{(\alpha k_B T)^2 + (2k_B T_{K})^{2}}$, 
being in this case $T_K = 43 \pm 1$\,K, with $\alpha$ an extra 
fitting parameter. 
This expression for the FWHM gives a value of $2k_B T_{K}/e$ at zero temperature, 
and is expected to coincide with the FWHM of the equilibrium spectral density 
for large asymmetric devices in the Kondo limit.\cite{capac} 
We must warn the reader that the Kondo scale determined from the width of the spectral
density can be 10\% larger that the corresponding one obtained from the
temperature dependence of the conductance.\cite{tksu2} The scaling of $G(V,T)$ for small 
$V$ and $T$ and also under an applied magnetic field has been
investigated experimentally and theoretically.\cite{grobis,kreti,scott,roura_3,ng,cb}

\begin{figure}[t]
\includegraphics[clip,width=6.0cm]{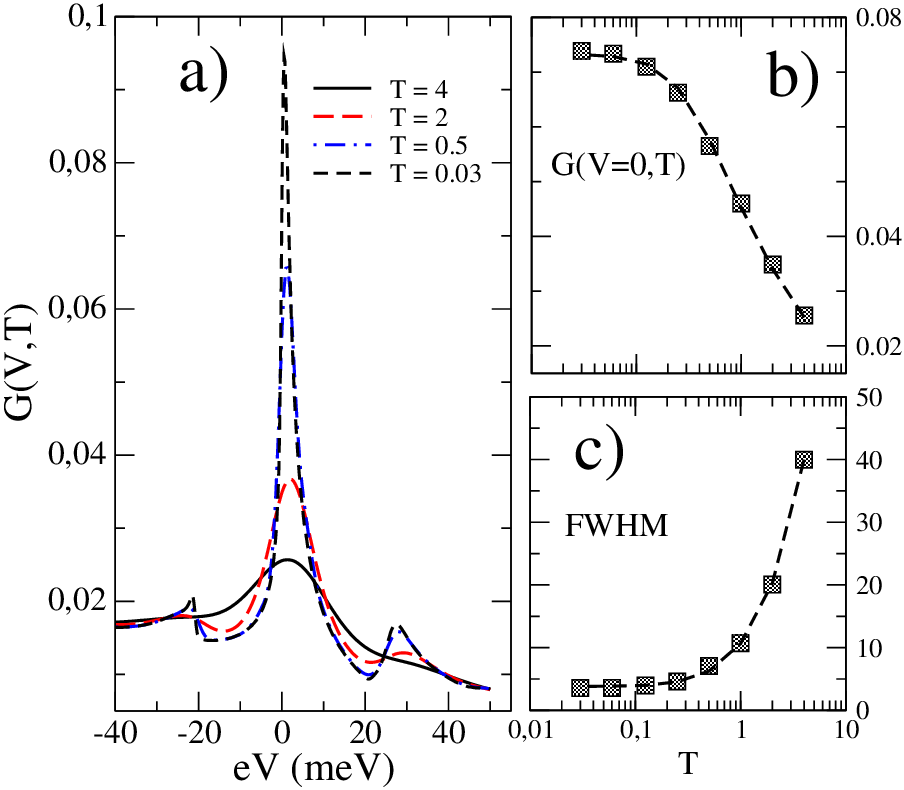}
\caption{(Color online) (a) Differential conductance
as a function of bias voltage for several temperatures and $\Omega=21$\,meV and $\lambda=12$\,meV. 
Other parameters as in Fig. \ref{fig-1}. (b) Equilibrium conductance $G(V=0,T)$ as a function of temperature (squares) and the scaling result using Eq.  (\ref{ley_empirica}) (dashed line). 
(c) FWHM of the zero bias peak as a function of temperature (squares) and the fitting result (dashed line).}
\label{fig-2}
\end{figure}

As we stated before, when the coupling to the vibration mode of the molecule is taken into account, 
it is not obvious that $G(T/T_K)$ and FWHM$(T)$ are still described with the same expressions. 
In Fig. \ref{fig-2}(a) we show the differential conductance as a function of the bias voltage for several 
temperatures and $\lambda=12$\,meV. All the parameters are the same as in Fig. \ref{fig-1}. 
As the temperature decreases, the zero-bias peak emerges together with lateral satellite peaks at voltages 
corresponding to the energy of the vibration mode $eV =\pm\hbar\Omega=\pm21$\,meV. 
Other satellites peaks with reduced intensity at multiples of the vibration energy 
should also be present,\cite{paaske,roura_1} but they are beyond the scope of this work.
The intensity of the two main satellite conductance peaks increases with increasing $\lambda$.\cite{paaske,roura_1} 
Fig. \ref{fig-2}b displays the zero-bias conductance as a function of temperature and the corresponding scaling 
result using Eq. (\ref{ley_empirica}). We obtained a Kondo temperature $T_{K}=20\pm 1$
K, which is markedly reduced due to the effect of the electron-phonon
coupling $\lambda $, but the decrease is smaller than expected from a
Franck-Condon factor. The non-trivial reduction of $T_{K}$ with the
electron-phonon coupling $\lambda $ has been discussed before.\cite{hm,mon,roura_1} 
We conclude that the only change in the temperature
evolution of the zero-bias Kondo peak is given by the Kondo scale, but the
universal behavior is not affected. 

The high precision of the fitting, with a correlation factor of $0.9995$,
indicates that for low enough temperatures and energies ($k_B T\ll\hbar \Omega$) the system 
behaves as a Fermi-liquid with renormalized parameters in which the phonon-mode is not active.

In Fig. \ref{fig-2}(c) we show the FWHM of the zero bias anomaly as a function of temperature 
and the corresponding scaling being $\alpha=9.9 \pm 0.5$ and 
$T_K = 22 \pm 1$\,K. This independent analysis supports the previous one obtained with the 
data from Fig. \ref{fig-2}(b)
and interestingly, the adjustable parameter $\alpha$ differs only by 4\% to the corresponding value 
without coupling to phonons (Fig. \ref{fig-1}(c). 

\section{The satellite peaks}

In what follows we focus on the analysis of the evolution with temperature of the two main satellite conductance 
peaks shown in Fig. \ref{fig-2}(a). In contrast to the left satellite, the maximum of the right one is slightly 
renormalized to higher voltages within the NCA. This might be due to the fact that the NCA does not incorporate 
renormalization of the bare phonon propagator, and it contributes in a different way for positive and negative 
frequencies.\cite{roura_1} However, we have verified that both side peaks
follow identical temperature dependence.\cite{note2}

\begin{figure}[t]
\includegraphics[clip,width=6.0cm]{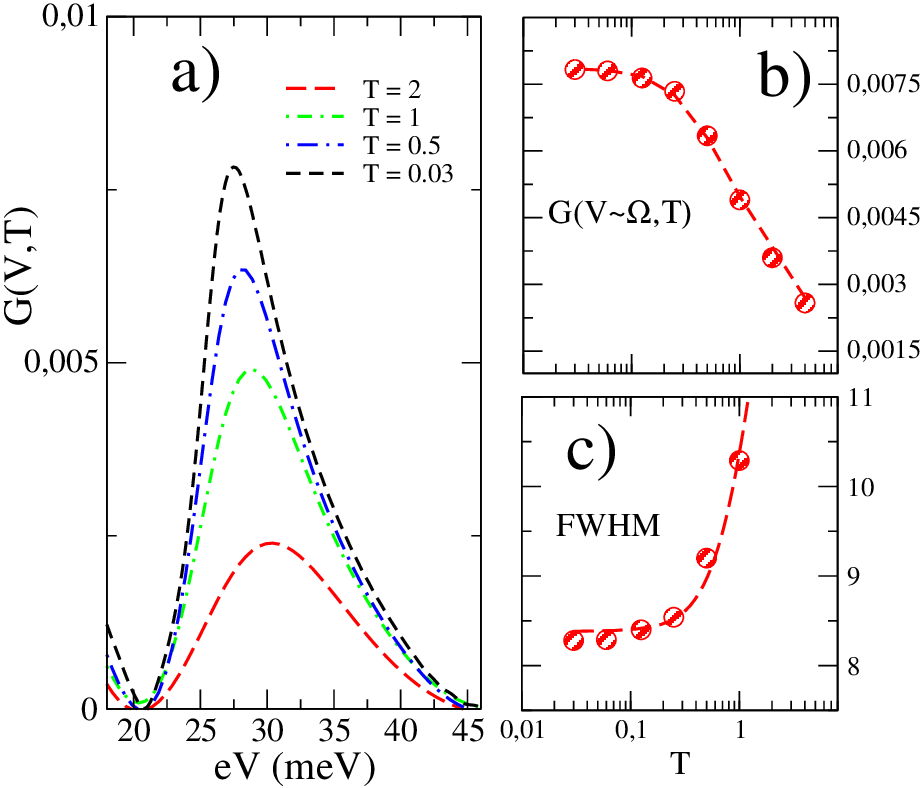}
\caption{(Color online) a) Differential conductance as a function of bias 
voltage for several temperatures. 
b) Differential conductance $G(V\sim\frac{\hbar}{e}\Omega,T)$ as a function of temperature (squares) and 
the corresponding scaling using Eq.  (\ref{ley_empirica}) (dashed line). 
c) FWHM (in meV) of the satellite peak at $V\sim\frac{\hbar}{e}\Omega$ as a function of temperature (squares) 
and the corresponding scaling (dashed line).}
\label{fig-3}
\end{figure}

In Fig. \ref{fig-3}(a), the satellite conductance at bias voltages near $eV\sim\frac{\hbar}{e}~\Omega$
is displayed after removing a linear offset for several temperatures.\cite{note4}  
From the maximum values, in Fig. \ref{fig-3}(b) we built the curve of $G(V\sim\frac{\hbar}{e}\Omega,T)$ as 
a function of temperature and its fitting by using Eq. (\ref{ley_empirica}). As in the experiment of 
Rakhmilevitch \textit{et al.}, the values of $G(V\sim\frac{\hbar}{e}\Omega,T)$ are very well represented
by the empirical law, being the adjustable parameter $T_K = 21 \pm 1$\,K in perfect agreement with
the Kondo temperature extracted from the central Kondo peak ($20$\,K).\cite{note5} 
Regarding the left satellite peak, we found $T_K = 24 \pm 1$\,K. 
On the other hand, we find that contrary to what happens for the Kondo peak, 
the width of the satellite peaks is not related to the Kondo temperature. 
Fig. \ref{fig-3}(c) shows the FWHM of the peak at $V\sim\hbar\Omega/e$ as a function of temperature. 
The parameters extracted from the fit differ with respect to those from the central peak, being $\alpha=6.1 \pm 0.5$ 
and a low-temperature width $49 \pm 5$\,K. The increase of the width is expected from the 
occurrence of inelastic processes. A similar behavior is found for the first satellite peak below the Fermi energy. 
We could not confirm that the same behavior is valid for further satellite peaks due to the lower
intensity of the latter and technical limitations of the NCA calculations.

Since the maxima of the conductance of the satellite peaks have the same temperature dependence as the 
Kondo peak, and are  scaled with almost the same Kondo temperature, we conclude that the side peaks are 
(broadened) replicas of the Kondo peak. 
As we stated in the introduction, this is not expected {\it a priori} and it would be desirable to have a physical 
explanation for this.
We have extended the variational approach of the Anderson model for the impurity spectral density below 
the Fermi level explained in Ref. \onlinecite{hewson} to include phonons. 
Performing perturbations at lowest order in the 
electron-phonon interaction $\lambda$, we find a replica at the expected position with relative 
intensity $(\lambda/\hbar\Omega)^2$. 
Basically, annihilation of the dot electron in the perturbed ground state leads 
in part (with an amplitude $\lambda/\hbar\Omega$) to the same excited states
as in the ordinary Anderson model except for the fact that they contain a phonon and thus their
energy is shifted by $\hbar \Omega$ (in leading order). This simple approach works qualitatively in  
the regime $\hbar\Omega \gg \textit{k}_B T_K$.
However for $\hbar\Omega \approx \textit{k}_B T_K$, the energy denominators in 
a more refined perturbative treatment might be near $\hbar\Omega \pm T_K$ and depending on the particular 
energy of the perturbed Kondo states, the amplitudes might be different, distorting the side peaks.
 
\begin{figure}[t]
\includegraphics[clip,width=6.0cm]{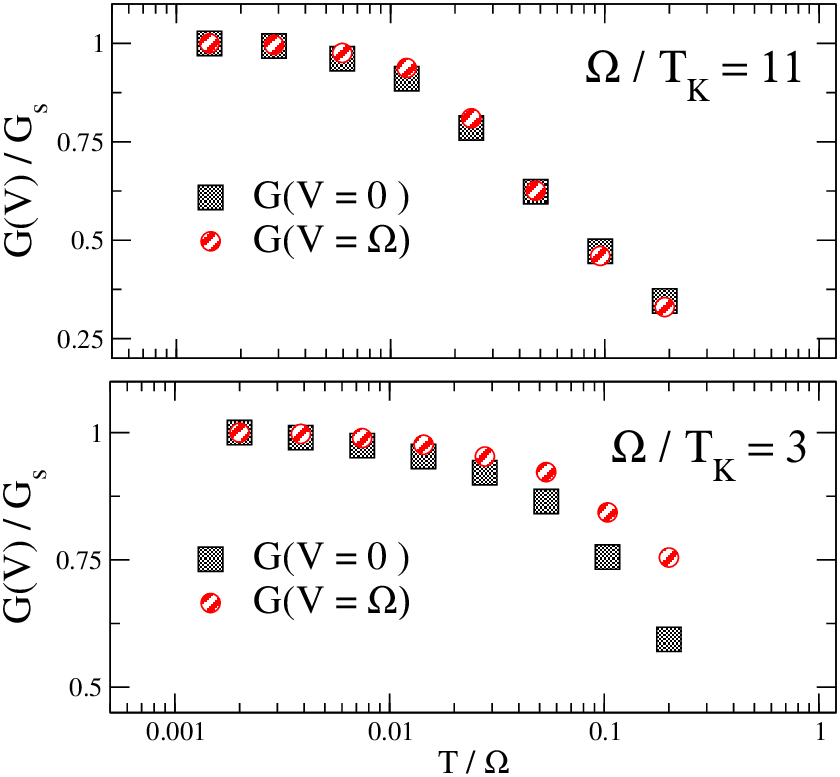}
\caption{(Color online) Top panel: Comparison of the normalized differential conductance, $G(V=0)/G_s$ and $G(V\sim\hbar\Omega/e)/G_s$, 
as a function of $k_BT/\hbar\Omega$ for the data in Fig. \ref{fig-2} 
and Fig. \ref{fig-3} respectively. Lower panel: same comparison as in the top panel for $\hbar\Omega=5$\,meV, $\Gamma_R+\Gamma_L=80$\,meV, 
$E_d=-40$\,meV, $\lambda=12$\,meV. The adjustable parameter from Eq. (\ref{ley_empirica}) is $19 \pm 1$\,K in the case of 
$G(V=0)/G_s$ and $30 \pm 1$\,K for $G(V\sim\hbar\Omega/e)/G_s$.}
\label{fig-4}
\end{figure}

In order to test the above physical picture, we study a different regime in which $\hbar\Omega$ is of the order 
of $k_B T_K$. For simplicity, we use a reduced value of the vibration frequency $\hbar\Omega=5$\,meV, changing 
the other parameters in order to analyze different values of the ratio $\hbar\Omega/k_B T_K$. The top panel of 
Fig. \ref{fig-4} displays a comparison of the normalized differential conductance, $G(V=0)/G_s$ and 
$G(V\sim\hbar\Omega/e)/G_s$, as a function of $k_BT/\hbar\Omega$ for the data in Fig. \ref{fig-2} 
and Fig. \ref{fig-3} respectively. As we already discussed, in this particular regime for which 
$\hbar\Omega/k_B T_K = 11$, 
the evolution with temperature of the zero-bias peak and the satellite conductance peaks is the same. 
On the other hand, the lower panel shows that the temperature evolution is no longer the same in the 
case of $\hbar\Omega/k_B T_K = 3$. Here we have used $\hbar\Omega=5$\,meV, $\Gamma_R+\Gamma_L=80$\,meV, 
$E_d=-40$\,meV and $\lambda=12$\,meV. For this choice of parameters the energy scale obtained from Eq. (\ref{ley_empirica}) is found to be
$19 \pm 1$\,K in the case of $G(V=0)/G_s$ and it still represents the Kondo temperature. 
In fact, this value agrees with the corresponding one obtained from the FWHM of the zero bias anomaly. 
However, the adjustable parameter extracted from a fitting of the temperature dependence of the 
inelastic peak at positive voltage $G(V\sim\hbar\Omega/e)/G_s$ is found to 
be $30 \pm 1$\,K, near twice larger than $T_K$. We have verified that this deviation is always present when 
$k_B T_K$ approaches $\hbar\Omega$
varying $\hbar\Omega$ from $5$ to $21$\,meV and also $E_d$ from $-40$ to $-90$\,meV. 
We cannot reach the range for which $\hbar\Omega/k_B T_K \sim 1$ due to the fact that larger values 
of $T_K$ drive the system towards a mixed valence regime for which there is no universal behavior of the conductance. 
A possible way to avoid this limitation would be to use unrealistic small values of the electron-phonon 
interaction $\lambda$. 

\section{Summary}

In conclusion, we have verified theoretically that the empirical expression for the temperature dependence 
of the conductance given by 
Eq. (\ref{ley_empirica}) still works to extract the Kondo scale in transport measurements through 
molecules with active phonon modes for temperatures  $\textit{k}_B T < \hbar\Omega$. It is also able to 
reproduce the temperature dependence of the out-of-equilibrium conductance peaks 
at finite bias voltages of the order of $\vert V\vert\sim\hbar\Omega/e$ with the same energy scale when 
$k_B T_K \ll \hbar\Omega$. These side peaks are however broader than the central Kondo peak due to inelastic 
effects. In cases for which $k_B T_K \sim \hbar\Omega$,  while the same expression is able to fit the 
temperature dependence of 
of $G(V=0)$ and $G(\vert V\vert \sim\hbar\Omega/e)$ the resulting energy scales do not coincide with each other, 
which is an indication of different functional dependence. As a concluding remark, we want to stress here 
that the ratio $\hbar\Omega/k_B T_K$ could be very different depending on the experimental system due to 
the particular frequencies of the vibration modes of the tested molecule and also due to the couplings of 
the molecule to the metallic contacts.

\section*{Acknowledgments}

We are partially supported by CONICET, Argentina. This work was sponsored by PICT 2013-1045 of the 
ANPCyT, Argentina, PIP 112-201101-00832 of CONICET, Argentina


\begin{thebibliography}{99}
\bibitem{note3} While the NCA cannot reach $T=0$, the conductance has
already reached its saturation value for temperatures $\sim 0.1 T_K$ for
which the NCA is still reliable.
\bibitem{comment_2} Other forms of this kind of expression, like $\frac{e^2}{
8k^2_B}(\mathrm{FWHM})^2 = (\pi T)^2 / 2 + T^{2}_{K}$, are also found in the
literature, \cite{rak,yu}.
\bibitem{note2} We have also replaced the bare phonon propagator by the
dressed one including corrections to second order in $\lambda $, resulting
in very small changes in the satellite peaks and to the same temperature
dependence.
\bibitem{note4} The satellite peaks are mounted on the tail of the main
Kondo peak (see Fig. 1). We have subtracted this tail assuming a linear
dependence, which is justified due ti the small curvature of the main peak
at the energies of the satellites.
\bibitem{note5} We note that if instead of following the maximum of the satellite peak
in $G(V)$ for each temperature, a fixed $V$ is taken (for example the position of the maximum
at $T=0$), the scaling law is not obeyed.
\end{thebibliography}
\end{document}